# Educating the Next Generation of Leading Scientists: Turning Ideas into Action

State of the Profession Position Paper
submitted to the Astro2010 Decadal Survey

Michael Wood-Vasey and Regina Schulte-Ladbeck (Pitt),
William Blair (JHU), Kirk Borne (George Mason), Mark Clampin (NASA/GSFC), Ian Gatley (RIT), Paul Graf (Aerospace Solutions, LLC), Zeljko Ivezic (U Washington), Eugene Magnier (IfA), John Mather (NASA/GSFC), Christopher Stubbs (Harvard), Andrea Schweitzer (Little Thompson Observatory), Tony Spadafora (LBNL), Phil Stahl (NASA), Tony Tyson (UC Davis)

**Our position**

The core of scientific research is turning new ideas into reality. From the school science fair to the search for the secrets of dark energy, high-quality research consists of scientific investigation constrained within the scope of a well-defined project. Large or small, generously funded or just scraping by, scientific projects use time, money, and information to turn ideas into plans, plans into action, and action into results. While we, as a community, do much to educate students in the techniques of research, we do not systematically train students in the nature and organization of scientific projects or in the techniques of project management. We propose a two-pronged attack to address this issue in the next decade. First, to generate a broad base of future scientists who have a basic familiarity with the ideas of projects, we propose that the community develop standards for the content of a project design and management course in astronomy and astrophysics. Second, to train future scientists to assume leadership roles in new investigations in astronomy and astrophysics, we propose that the community develop standards for graduate programs in the area of research project leadership.

**Statement of the Problem**

Astronomical research is increasingly done in sponsored projects. Today's successful astronomy professor is as much a proficient project manager as a proficient scientist and teacher. Moreover, professional astronomers work outside academia – in business, government, and non-profit organizations – where project management skills have widespread application. A serious shortcoming of astronomy education is that we do not teach our students the project management skills they need for both research and later careers.

We suggest that the following topics should become an intrinsic part of undergraduate and graduate education throughout the STEM area, and astronomy in particular.

1. Project management, both experiential and theoretical.
2. Proposal writing, both as a tool for obtaining support and as an example of a project.
3. Presentation skills, using the proposal as an example of winning support for an idea.
4. Research strategy and experimental design, taking a large objective and breaking it down into manageable pieces that work individually and together.

In this paper, we outline a suggested set of requirements and an approach to meeting them that we recommend to those designing curricula for astronomy students. Our key point is that there ought to be standards that we as a community agree upon.

**Introduction**

If you ask the typical astronomy graduate student to come up with a new research project—be it theoretical, observational, or experimental in nature—involve in it the right people to maximize available expertise, and finish it in three years, that student will likely look at you blankly and have little idea of how to proceed. We consider the postdoctoral years as the time of maturation and an eventual introduction to research leadership and the induction into professional roles. But in reality we teach postdocs no more than graduate students about how to actually develop a vision for scientific investigation, and then realize it.

New ideas are key to moving science forward. Productive astronomers have the ability to envision and recognize the best ideas and they effectively implement research strategies to tackle them. However, designing, managing, and bringing a scientific research project to completion requires more than original scientific thinking; it also requires project management skills which are well understood, but are currently not well taught in our graduate programs.

Many new great ideas fail to gain support because the proposers have not thought out how the project will actually be done. Often one has an inescapable feeling that the proposers do not really know how to organize and implement their idea. This is one of the reasons that past performance has such influence in the proposal approval process. Those who have demonstrated that they can lead successful scientific projects have a significant advantage when applying for subsequent funding. This is in general a good criterion, but in order to continue to encourage new ideas from new people, we must ensure that new astronomers are capable of turning their ideas into well-thought-out proposals with plans for likely success.

We often despair at the convoluted paths our undergraduate students take to arrive at a solution to a problem. And this is generally for well-posed problems with well-defined answers. We teach students to be systematic. We teach them skills to break down problems such as the free-body diagram, the right-hand rule, or the harmonic oscillator.

However, if you look closely, the same phenomenon is observable in how graduate students approach research. They did not magically learn how to initiate, design, and execute a research project in the summer after college. We celebrate and inwardly delight at those graduate students who start their education ready to propose and tackle new experiments, and knowing how to garner the requisite expertise, but this is precisely because this ability is not the norm. We can and should explicitly teach these skills.

We believe that we have an opportunity to improve the profession by teaching and training the majority of astronomy students to take ideas from imagination to practice.

We believe that enhancing science education in the U.S. to provide training in leadership, and research-project design and management in the next decade, will quickly repay a modest investment by the community, through improving the effectiveness and productivity of astronomical projects.

We believe the proposed plan outlined here will increase the benefits not just to astronomy but to the majority of Ph.D.s who will work outside the traditional academic sphere.

**What is a project?**

In the most generic sense a project is a temporary endeavor undertaken to create a unique product, service, or result (Project Management Institute, 2004). It has a scope, includes a start and finish within an identified period of time, and requires resources. Put broadly most scientific investigations fall under this definition of project.

Framing research as projects is informed by a pragmatic view to scientific inquiry, whereby the scope and lifecycle of an investigation are adjusted to the needs of and the resources provided by society. Thinking of research in terms of projects has the advantage of continually testing progress against a set of objectives. This approach enables project-bound researchers to abandon failed lines of inquiry and, for the time being, move on to more promising ones, at an earlier stage of the investigation.

**Managing small and large projects**

Many research projects in astronomy and astrophysics are small- to medium-sized collaborative enterprises that involve, on average, a half-a-dozen astronomers with a variety of expertise (Abt, 2000, updated by a search on the number of authors per paper in 2008 using the NASA Astrophysics Data System). The researchers' efforts are usually internally coordinated by one member of the collaboration who often also serves as the first author of the resulting publication. In the U.S., many research endeavors are sponsored projects that use public monies distributed by federal agencies. Here, the principal investigator (PI) of the collaboration has the overall responsibility for carrying out the project and is accountable to the sponsor for project success. Every scientist who has undertaken a sponsored research project has experienced administrative and leadership tasks, such as planning and budgeting, fundraising, distributing tasks, motivating and coordinating staff, and writing reports.

A very practical distinction between small and large projects is that small projects can get by without realizing that they are projects while large projects will fail miserably without organization, planning, and orchestrated execution.

 "Big Science" projects (Weinberg, 1961) have become the hallmark of the natural sciences in the last century and often necessitate decades of leadership to realize. Astronomy has its share of Big Science projects. Many are global enterprises. The Hubble Space Telescope, for example, required thousands of scientists and engineers to design, build, launch, operate, use and eventually de-orbit; and as of 2008 had cost approximately ten billion dollars (Moskowitz, 2008). Without doubt, the Astro2010 decadal survey will endorse new large facilities and big projects that our community will create in the next decade and beyond.

A few more examples of large, recent, U.S.-led projects in astronomy and astrophysics are the James Webb Space Telescope, the Sloan Digital Sky Survey, the Hubble Deep Field, the Via Lactea Project, the U.S. International Year of Astronomy 2009, Google Sky and of course this very Astro2010 decadal survey. One common aspect of all of these projects is that they draw on the creative power of a diverse multitude of scientists to explore a vast array of ideas in groups large and small. Another aspect is that information products are being shared internally and made publicly available for external users.

Large research projects can be thought of as virtual communal networks that connect smaller research nodes across the globe to a hub – a core project team or telescope facility or national lab or data repository. An effective strategy for sponsors of large projects is to draw on world experts, use the de-centralized expertise that exists in independent medium- or small-sized research communities, and integrate them into the whole. A medium-sized research group might involve several senior scientists, a couple of energetic young faculty, a few postdocs, and a half-dozen graduate students and may have the know-how to build the world's most efficient detectors. A small research group might consist of a professor, a graduate student, a colleague at another institution, and a couple of undergraduates and may have developed the code that best models the astrophysics involved in galaxy evolution. These smaller groups will be working on smaller projects of their own and will develop their own organizational expertise while at the same time being partners in larger collaborative scientific endeavors and driving communal expertise.

Astronomical projects and collaborations exist on a wide range of scales, and they are often interconnected to stimulate new ideas and share expertise. Organization furthers the effective use of originality. The organized sharing of thoughts, plans, and ideas provides the rich fertile ground for creative thinking. The challenge is how to marshal the efforts of a team to come up with new ideas and creative efforts either in the context of a small research group or as a sub-group of a large project. Organizing the connection of a smaller group to the large whole, whether that be the community at large or a larger collaboration, is key in allowing ideas to blossom and new ideas to spread. Just as the prepared create their own luck, it is the organized team that can quickly respond to challenges and adapt to new information and discoveries.

**What is science leadership?**

Scientists provide leadership in the generation of new knowledge and in the use of societal resources that are needed for discovery. To reduce it to a quantifiable basis, science leadership can be characterized as having two dimensions, originality and stewardship (Schulte-Ladbeck, 2009). Originality is about new ideas. It ranks a scientist's intellectual impact on science. Stewardship is about efficiently guiding a project to success; it measures how a scientist uses public resources to get things done in collaboration with other scientists. Figure 1 illustrates the taxonomy of science leadership, with scales arbitrarily ranging from 1 to 10. Each corner represents an idealized extreme of science leadership, which we dub leading scientist, science leader, and science champion.

The "leading scientist" generates new concepts. This scientist ranks high in originality, but in as much as s/he does not use societal resources that lead from thoughts to new observations about the Universe, the leading scientist rates low on stewardship. When a scientist leads by being the catalyst for new thoughts and acts on them to create the conditions that turn them into certified knowledge, we call this role a "science leader." A science leader articulates a relevant question and finds the means to answer it. This role includes leading a team of experts and providing them with the resources they need to do their part in the investigation. Like the leading scientist, so also is the science leader recognized for individual original achievement. The community leaves it up to him or her to share recognition and rewards by allocating credit to the team. Scientists also employ their creativity to provide other scientists with the means to do original work. The "science champion" enables others to get science done. A scientist in this role leads in service to research projects and their principal investigators. In Schulte-Ladbeck (2009) we had called this role a science administrator; however, because there are many ways in which scientists make creative contributions that are critical to the success of research projects, such as by designing new instrumentation or developing new computer code, we find science champion a more inclusive descriptor. While the science champion's contribution to science is critical, it does not result in first authorship on discovery papers. Because the reward system in science is focused almost exclusively on originality, scientists who excel at pure stewardship roles are not usually afforded the same recognition and high status in the community as leading scientists and science leaders.

Over the course of their lives, scientists' roles change and can become more akin to the leading scientist like, science leader, or science champion role. Some scientists experience a sequence of such roles throughout their careers. Their lives can be charted as tracks on this diagram, with changes in direction occurring over time by individual career choice, chance, or other influences from the scientist's environment.

It is our position that the profession has a need for leading scientists, science leaders, and science champions alike. Instead of migrations on Fig. 1 induced by age and life experience, the profession should strive to actively select and educate scientists for these different roles. This strategy will require a combined effort between funding agencies to specifically support grants for personnel in these areas, institutions to provide attractive and permanent positions to these traditionally less-supported roles that are increasingly important in enabling astronomical science, and degree programs to educate their students about these possibilities. In this paper we focus on preparing the next generation of students by teaching them about each of these two basis functions and equipping them to choose and understand their journey and interactions along this plot.

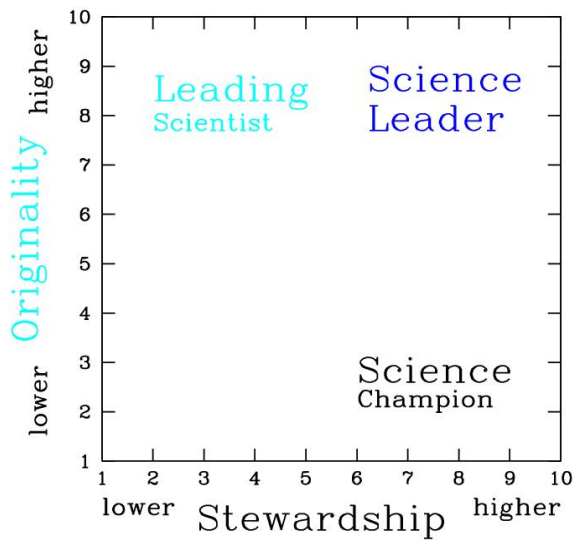

**Figure 1**. Taxonomy of science leadership (after Schulte-Ladbeck, 2009). While the entire plane is populated by scientists, there are idealized roles that are defined by extreme values of originality and stewardship. Over the course of their lives, scientists tend to experience one, or more than one, job that approximate these roles.

**Why should we teach project leadership?**

1. *The culture of science has changed.*

Astronomers, like other scientists, value individual, high achievement. That is why we push ourselves and our students to make that leapfrog original contribution and to become one of the leading minds of science. The holy grail of achievement is to have a theory, or perhaps a law, or maybe a scaling relation named after us, as the everlasting legacy of our contribution. We romanticize and admire the lonely genius, particular the one who eschews the standards of his or her day, believing in self-sufficiency and an innate ability to master any new task or challenge.

> "**Management training? Why do I need to go to management training? I have a Ph.D. in physics.**" As quoted in Clarke (2002).

The ranking of a scientist along the originality axis will continue to be important to us. Yet, many of our endeavors today are collaborative in nature, and approach in size the typical particle physics collaboration. Take, for example, the recent generation of survey projects, such as the Sloan Digital Sky Survey or the Large Synoptic Survey Telescope. They operate under the paradigm of collective efforts of many scientist, engineers, and IT professionals, toward a data product that is accessible and useful to the entire astronomical community. Our education is now too narrowly focused on advancing scientists along the originality axis. "Big Science is here to stay, but we have yet to make the hard [] educational choices it imposes," wrote Weinberg in his seminal 1961 article. It is long overdue for us to consider how we should modify university education to enable the next generation of scientists to operate and to lead effectively in increasing larger, more international, and more collaborative settings. This will strengthen us as a community, and it will better prepare us to avoid such catastrophic failures as the demise of the Superconducting Super Collider, a major set-back to U.S. leadership in particle physics (Riordan, 2000).

2. *The careers of scientists have changed.*

Many of us are already experiencing double (research and research management) and even triple (commercialization management) career paths within academia. In almost all cases, we learned how to conduct our non-science jobs by experience. This approach is ineffective. We all make the same mistakes as we learn on our own instead of benefiting from established knowledge.

Some of us had Ph.D. advisors who modeled their leadership behaviors or passed on their experiences to us. They may even have helped us to conceive of our Ph.D. thesis as a project, with a finite duration, a limited amount of funding support, and a scope. However, there is much variation in the quality of faculty-student relationships, with the result that some of them do not significantly further the professional training or career development of the student.

Project leadership in particular stands out as an activity that is an important part of the work life for most Ph.D. scientists, while only very few receive significant training (including informal training) in this area (Smith, Pedersen-Gallegos, and Riegle-Crumb, 2002). This finding emphasizes the impact that a well thought-out formal management and leadership curriculum could have on science professions. Making leadership and project management a part of graduate education in the sciences will both build capacity for future science leaders and champions and will shorten their experiential learning cycle.

> Davis (2006) blogs: "**We as a community don't do a lot to prepare younger scientists to take on leadership roles. [...] The attitude I sensed while I was an academic was that leadership activities were not for younger people. Younger people should keep focused on science. Only when you get tenure, when your work slows down, when you can't do science 24x7, should you engage in such things. To see the problem with this approach, turn it around. What kind of science would we have if there was no training and people were discouraged from engaging in any real research activities until they were well into their 50's?**"

We continue to educate our undergraduate students toward the goal of becoming graduate students, and our graduate students to become the next generation of the professoriate. Yet, it is only a small minority of our students who will find work in the academy. The majority of our graduates will become technicians or scientists working for businesses, government agencies, or non-profit organizations. In all of these "real world" settings, project management and leadership skills are paramount. It is timely that we, as stewards of the profession, adopt a more pragmatic outlook on education and train astronomers and astrophysicists for a wide range of career paths. Originality and stewardship together are needed by scientists to fulfill leadership roles in science and in society.

**How should we teach project leadership?**

Our emphasis is on teaching the next generation of students about leadership and project management in the context of how we do research in scientific collaborations. We advocate a two-pronged approach.

1. *Develop cognizance.*

To increase the capacity of astronomers who might become science leaders or science champions, we make the case that the graduate curriculum of the future must include at least one project-based leadership and project management course.

Here we outline what we consider the minimal standard for student learning outcomes of such a course. We also provide a course description with proposed pedagogy and a list of lecture topics. The exact nature of the research projects and lecture topics will vary depending on the university, its research facilities, and the specialization of its faculty. We believe that the course delivery could be adapted to suit students at both the undergraduate and the graduate level.

The study by Smith, Pederson-Gallegos, and Riegle-Crumb (2002) showed that Ph.D. physical scientists characterize their work as requiring communication, analytic, critical-thinking, and management skills in an interdisciplinary context. It may therefore be prudent for astronomers to form partnerships with science, engineering, and business faculty for the development and teaching of a joint course for graduate students from all science departments.

Because the project management profession has a globally agreed-upon knowledge and skill set, and while scientists do manage projects, as instructors of this course science faculty might first need to obtain instruction in professional project management themselves.

> **Strawman Course "Leadership and Project Management for Scientists"**
>
> *Standards for student learning*
> After completing this course, students will be able to:
> - describe strategies for designing a research project
> - recount the knowledge and skills needed for managing a research project
> - apply project management tools and techniques to a simple research project
> - collaborate in a project team
> - communicate about a research project orally and in writing
> - describe leadership styles for leading research projects and personnel.
>
> *Course description*
> This course has both an experiential and a theoretical component. To gain project experience, the course will be centered around research projects. Each student will come up with an idea for a research project for which s/he will be the principal investigator, or PI, and will collaborate on at least two additional projects. PIs must involve at least two other students from the class in developing a grant proposal; individual proposals will not be accepted. At midterm, all PIs must hand in a written proposal with a budget and a 6-week work plan, and give an oral presentation explaining the proposal. The instructor will rank all proposals and select a subset for funding. The sponsored PIs will implement their research projects and select other students to join their teams. At the end of the term, each team must submit a written final project report, and the PI must give an oral presentation of their research paper. In parallel to project activities, lectures will introduce students to theoretical concepts relevant to science leadership and project management. Lecture topics will address (1) the nature of science (reward system; structure of collaborations; resource availability; big science projects; science and society); (2) research skills (designing research projects; scientific ethics and integrity; reading, writing, and presenting papers); (3) project management knowledge and techniques (project management methodology; balancing project constraints; project roles and responsibilities; scope statement; work breakdown structure; estimating, scheduling, and budgeting methods and tools; status reports; scalability of project tools and techniques); (4) leadership skills (project leadership; the managerial grid; participative leadership; team leadership; leading in different cultures; leading virtual teams). Course grades will be based 50% on the ranking of students' individual research proposals; 25% on the teams' final project reports; and 25% on sponsored PI's oral presentations. All team members including the PI will receive the same score.

2. *Develop competence.*

Students who discover an aptitude for leadership and project management need to be given the opportunity to study a broader "science plus" graduate curriculum. Now is a good time to leverage professional education for future astronomers. The Professional Science Master's degree (PSM), whose development at U.S. universities was seeded in 1997 by funding from the Sloan Foundation (National Professional Science Master's Association, 2008) is receiving renewed attention and financial support. The National Academy of Sciences recently published a study that calls for an acceleration and spread in the development of professional type science master's education to meet the growing national need for science professionals (Board on Higher Education and Workforce, 2008). The goal is to prepare professionals who bring scientific knowledge *plus* "the ability to anticipate, adapt, learn, and lead where and when needed in industry, government and non-profits." The PSM has been included in the stimulus bill known as The American Recovery and Reinvestment Act of 2009. Accordingly, the National Science Foundation will receive up to fifteen million dollars for PSM Programs (Joint Explanatory Statement of Division A, 2009).

To meet the needs of the U.S. knowledge economy as well as that of future astronomers, we propose that universities graft and implement PSM degrees in astronomy and astrophysics. There are about 130 PSM programs that have produced 2100+ graduates across scientific disciplines. The list of PSM programs includes 14 in physics and geosciences (http://www.sciencemasters.com/PSMProgramList/ProgramsbyField/tabid/81/Default.aspx), but none focus on astronomy. The details of a "PSM in Astronomy" will vary from institution to institution, based on its research facilities and the interests of its faculty. An important aspect to consider is possible partnerships with local business, government, or non-profit organizations that would offer internships to students and eventually hire the graduates of the program.

Guidelines for PSM programs by the Council of Graduate Schools are available online at http://books.nap.edu/openbook.php?record_id=12064&page=37. We have several general comments about the program goals and standards for student learning. The PSM degree could be useful beyond its stated purpose if curricula were designed as a stepping stone for Ph.D. candidates interested in a more professional education. The "plus" component of the degree would depend largely on how academic and non-academic partners envision transfer of academic knowledge and skill into the workplace, i.e., with more of a research, development, service, or production orientation. Standards for student learning in physics and astronomy need to be at the level of knowledge and problem-solving skill appropriate for the master's degree. Mastery of the project management body of knowledge needs to be at the level appropriate for certification as project management professionals.

**Recommendations**

We make the following recommendations to the State of the Profession Group on Education and Public Outreach:

1. To generate a broad base of future scientists who have a basic familiarity with the ideas of projects, colleges and universities should modify their science curricula to include a required course in project management and leadership.

2. Because the teaching of project management knowledge and skills must become an integral part of teaching science, the professional societies – such as the American Astronomical Society – should call on the professional project managers among their members to provide "teaching the teachers" opportunities for colleagues on the science faculty.

3. To respond to the national need for a stronger scientific workforce, universities that already have, or have the potential for, strong regional partnerships with local businesses, government, or non-profit organizations, should develop professional graduate tracks that can culminate in a Professional Science Masters degree.

4. Because the effective stewardship of public resources benefits both science and society, colleges and universities should include stewardship as an explicit criterion for hiring and promotion decisions.

## Acknowledgments

We thank our generous colleagues for the many productive and stimulating discussions we have had in the preparation of this paper. Some ideas for the pedagogy of the strawman course came from Dr. Marai's course at the University of Pittsburgh (http://vis.cs.pitt.edu/teaching/cs2620/).